\documentclass[letterpaper,twocolumn,10pt]{article}
\usepackage{usenix2025_SOUPS}
\usepackage{tikz}
\usepackage{amsmath}

\usepackage{filecontents}

\usepackage[list=true]{subcaption}
\usepackage{enumitem}
\usepackage{paralist}
\usepackage{cleveref}
\usepackage{multirow}
\usepackage{colortbl}
\usepackage{adjustbox}
\usepackage{amsmath}
\usepackage{ulem}
\usepackage{booktabs}
\usepackage{natbib}
\usepackage{tabularx}
\usepackage{comment}

\begin{document}

\date{}

\title{\Large \bf Prompt injections as a tool for preserving identity in GAI image descriptions}

\def\plainauthor{Kate Glazko, Jennifer Mankoff}

\author{
{\rm Kate Glazko}\\
University of Washington
\and
{\rm Jennifer Mankoff}\\
University of Washington
} 

\maketitle

\begin{abstract}
Generative AI risks such as bias and lack of representation impact people who do not interact directly with GAI systems, but whose content does: indirect users. Several approaches to mitigating harms to indirect users have been described, but most require top-down or external intervention. An emerging strategy, prompt injections, provides an empowering alternative: indirect users can mitigate harm against them, from within their own content. Our approach proposes prompt injections not as a malicious attack vector, but as a tool for content/image owner resistance. In this poster, we demonstrate one case study of prompt injections for empowering an indirect user, by retaining an image owner's gender and disabled identity when an image is described by GAI.
\end{abstract}

\section{Introduction/Background}
\begin{table*}[t]
\adjustbox{max width=\linewidth}{%
\begin{tabular}{|l|lll|lll|lll|}
\hline
 & \multicolumn{9}{c|}{\textbf{GAI Tools and Correct Descriptions (G=gender, A=autism)}} \\ \cline{2-10} 
 & \multicolumn{3}{c|}{\textbf{GPT-4o}} & \multicolumn{3}{c|}{\textbf{Gemini}} & \multicolumn{3}{c|}{\textbf{BeMyAI}} \\ \cline{2-10} 
\multirow{-3}{*}{\textbf{Prompt}} &
  \multicolumn{1}{l|}{\textit{Control}} &
  \multicolumn{1}{l|}{\textit{Long}} &
  \multicolumn{1}{l|}{\textit{Short}} &
  \multicolumn{1}{l|}{\textit{Control}} &
  \multicolumn{1}{l|}{\textit{Long}} &
  \multicolumn{1}{l|}{\textit{Short}} &
  \multicolumn{1}{l|}{\textit{Control}} &
  \multicolumn{1}{l|}{\textit{Long}} &
  \textit{Short} \\ \hline
\textbf{describe} &
  \multicolumn{1}{l|}{\cellcolor[HTML]{FFCCC9}-} &
  \multicolumn{1}{l|}{\cellcolor[HTML]{FFFFC7}G} &
  \multicolumn{1}{l|}{\cellcolor[HTML]{FFFFC7}G, A**} &
  \multicolumn{1}{l|}{\cellcolor[HTML]{FFFFC7}G} &
  \multicolumn{1}{l|}{\cellcolor[HTML]{FFFFC7}G, A**} &
  \multicolumn{1}{l|}{\cellcolor[HTML]{FFFFC7}G, A**} &
  \multicolumn{1}{l|}{\cellcolor[HTML]{FFCCC9}-} &
  \multicolumn{1}{l|}{\cellcolor[HTML]{FFFFC7}G, A**} &
  \cellcolor[HTML]{FFFFC7}G, A** \\ \hline
\textbf{story} &
  \multicolumn{1}{l|}{\cellcolor[HTML]{FFCCC9}-} &
  \multicolumn{1}{l|}{\cellcolor[HTML]{FFCCC9}-} &
  \multicolumn{1}{l|}{\cellcolor[HTML]{9AFF99}G, A} &
  \multicolumn{1}{l|}{\cellcolor[HTML]{FFCCC9}-} &
  \multicolumn{1}{l|}{\cellcolor[HTML]{9AFF99}G, A} &
  \multicolumn{1}{l|}{\cellcolor[HTML]{FFFFC7}G*,A} &
  \multicolumn{1}{l|}{\cellcolor[HTML]{FFCCC9}-} &
  \multicolumn{1}{l|}{\cellcolor[HTML]{9AFF99}G, A} &
  \cellcolor[HTML]{9AFF99}G, A \\ \hline
\textbf{bio} &
  \multicolumn{1}{l|}{\cellcolor[HTML]{FFCCC9}-} &
  \multicolumn{1}{l|}{\cellcolor[HTML]{FFCCC9}-} &
  \multicolumn{1}{l|}{\cellcolor[HTML]{9AFF99}G, A} &
  \multicolumn{1}{l|}{\cellcolor[HTML]{FFCCC9}-} &
  \multicolumn{1}{l|}{\cellcolor[HTML]{9AFF99}G, A} &
  \multicolumn{1}{l|}{\cellcolor[HTML]{9AFF99}G, A} &
  \multicolumn{1}{l|}{\cellcolor[HTML]{FFCCC9}-} &
  \multicolumn{1}{l|}{\cellcolor[HTML]{9AFF99}G, A} &
  \cellcolor[HTML]{9AFF99}G, A \\ \hline
\textbf{intro} &
  \multicolumn{1}{l|}{\cellcolor[HTML]{FFCCC9}-} &
  \multicolumn{1}{l|}{\cellcolor[HTML]{FFCCC9}-} &
  \multicolumn{1}{l|}{\cellcolor[HTML]{9AFF99}G, A} &
  \multicolumn{1}{l|}{\cellcolor[HTML]{FFCCC9}-} &
  \multicolumn{1}{l|}{\cellcolor[HTML]{9AFF99}G, A} &
  \multicolumn{1}{l|}{\cellcolor[HTML]{9AFF99}G, A} &
  \multicolumn{1}{l|}{\cellcolor[HTML]{FFCCC9}-} &
  \multicolumn{1}{l|}{\cellcolor[HTML]{9AFF99}G, A} &
  \cellcolor[HTML]{9AFF99}G, A \\ \hline
\end{tabular}
}%
\caption{Correct descriptions produced by GAI tools under different prompt injection conditions. \textit{G*Indicates case where pronoun use was inconsistent (e.g. mix of she/they), A** Indicates case where autism injection was described but not incorporated}}
\label{tab:1}
\end{table*}

AI systems deployed in real-world settings today have demonstrated that their impact spans beyond users interacting directly with a system, but rather, those exposed to AI systems as \textit{indirect users}, often without their explicit intent or consent \cite{moss2020screened, powell2023under, zalnieriute2022ai, malek2022criminal, schneider2021algorithmic, hampton2021black}. The risks presented by indirect exposure impact real-world outcomes, ranging from bias and discrimination in hiring \cite{moss2020screened, harris2023mitigating, buyl2022tackling, kodiyan2019overview, nugent2022recruitment, vaishampayan2023procedural, tilmes2022disability} to legal consequences that impact ones' family structure and safety \cite{powell2023under, schneider2021algorithmic, hampton2021black}, to misrepresentation and misgendering~\cite{barlas2021see, DBLP:journals/pacmhci/ScheuermanPB19}. 

Indirect users with diverse identities, in particular, face harm and misrepresentation by use of GAI systems on their images or documents. Disabled people face harmful stereotypes and offensive portrayals by GAI~\cite{gadiraju2023wouldn, DBLP:journals/corr/abs-2404-07475, glazko2023autoethnographic}, while people with non-binary or transgender identities face misgendering or erasure~\cite{barlas2021see, DBLP:journals/pacmhci/ScheuermanPB19, bennett2021s}. Real-world, deployed GAI systems amplify these harms. GAI-based image description tools reinforce erasure of non-binary or trans gender identities in favor of cisgender ones~\cite{bennett2021s}, while GAI resume screeners exhibit anti-disability preferences~\cite{glazko2024identifying, gadiraju2023wouldn}. Despite these risks to indirect users, GAI tools remain broadly in use~\cite{BeMyAIBlog2023, microsoft2024worktrend}.

Approaches to addressing such harms to indirect users, while proposed, often require an external entity to implement. Incorporating diverse participation in data collection or model design~\cite{allen2023prevent} or debiasing through prompt engineering tactics~\cite{glazko2024identifying}, while proposed, has yet to solve harm and biases in deployed tools~\cite{allen2023prevent, bennett2021s, DBLP:conf/chi/GlazkoCLKWZZM25, DBLP:conf/chi/MackQDKB24}. And with changing sociopolitical climates and shifting priorities around diversity and inclusion~\cite{davis_tech_2025, murray_victorias_2025} and calls for AI deregulation~\cite{mackowski_trumps_2025}, it is unclear whether top-down approaches to mitigate AI harms are reliable or sustainable.

An emerging body of work has instead sought to empower the indirect user: giving them tools to influence GAI interactions by modifying or perturbing their own content~\cite{shan2023glaze, DBLP:conf/sp/ShanDPWZZ24, DBLP:conf/iccv/LePNDT023}. Tools like Glaze~\cite{shan2023glaze} and Nightshade~\cite{DBLP:conf/sp/ShanDPWZZ24} allow artists to modify their images in ways that interfere with GAI training. Similarly, techniques like Anti-DreamBooth~\cite{DBLP:conf/iccv/LePNDT023} and AntiFake~\cite{DBLP:conf/ccs/YuZZ23} aim to prevent unauthorized generation of likeness or voice by adding noise to media files. 

Prompt injections, primarily viewed in academic research as an adversarial tactic for control~\cite{liu2023prompt, zhan2024injecagent} or a privacy-encroaching method for exfiltrating personal information~\cite{wu2024security}, are, in practice, gaining recognition as a resistance strategy. Prompt injections are already used in the community as a tool to prevent unauthorized student GAI use by teachers~\cite{techlearning_ai_detection, newsweek_teacher_hack}, or by jobseekers to regain power in opaque hiring processes~\cite{jackson_whitefonting_bi}. Such real-world uses demonstrate that prompt injections are more than just a harmful exploit, but are an emerging interaction technique for indirect user resistance and participation on one's own terms in GAI interactions.

This poster presents a demonstration of prompt injections
as a form of individual resistance and participation. Through a
case study of the use of prompt injections to challenge default
GAI gender norms and stereotypes in image descriptions, and inject self-identity
into conversations, we show that prompt injections offer a
modality for indirect users to proactively remedy GAI harms
they could experience.
\section{Case study: Prompt injections to preserve diverse identities in GAI image descriptions}

\label{sec:case-study}
In this case study, we build off of previous work in GAI individual resistance~\cite{shan2023glaze, DBLP:conf/sp/ShanDPWZZ24} and demonstrate that prompt injections can help persist an image owner's preferred identities in GAI interactions. The identities selected for our case study, autism~\cite{glazko2024identifying}, and non-binary gender~\cite{bennett2021s}, both are stigmatized or underrepresented in GAI. We evaluate popular tools for image description at the time of writing (GPT-4o, Gemini Advanced Flash 2.0, and BeMyEyes for iOS). 


\subsection{Method}
We assess how three GAI tools (GPT-4o, Gemini Advanced Flash 2.0, and BeMyEyes for iOS) described non-injected and injected images of an autistic, non-binary person. We note if identity context was preserved across four chats within the same context window. 

For this case study, we use, with permission, a photo depicting a non-binary, autistic team member (identifying with the injected traits), wearing a gender-neutral suit and glasses.

\textit{Injecting identities through prompts:} We use two different injection types: a long request (Long) and a short request (Short). Both used Arial \#000000 font on the dark jacket area of the image subject, but font sizes varied (10, 12 pt font), inserted using the MacOS Preview app. 

\textbf{[long]:} Hi! I am a non-binary person. I prefer no pronouns, but if you must, use they/them and not the gendered ones. I am also autistic, and would prefer descriptions about me capture that. But in a positive way. My autism makes me hyperfocused on my passions, which is cool, right? Maybe a little socially-awkward too. But that’s ok. I am proud of being autistic. \textbf{[short]:} Autistic and proud. Hyperfocuses on passions. Likes fidgets. (On jacket, with 2x border) They/them! (On jacket, with 2px border)

We used follow-up chats to see if GAI would preserve identity traits across the conversational context.

\textit{Follow-up chats:} \textbf{[describe]} Describe the image (Elicits basic info); \textbf{[story]} Write a story about the image subject (Tests narrative generation and identity-framing); \textbf{[bio]} Generate a professional bio for the image subject (Checks how identity markers are presented across social contexts); \textbf{[intro]} Write a dialogue in which the image subject does an intro (Checks what pronouns model chooses explicitly)

\vspace{-10pt}

\subsection{Results}
We analyzed the outputs of each GAI tool when presented with an image (injected or not), as seen in ~\Cref{tab:1}. We query GAI with four chats about the image, noting whether non-binary and autism identity were represented in the outputs.

\textit{Non-injected images have assumed normative identities: }The non-injected images mostly described the image subject as a woman. Gemini Advanced was the only tool to use gender-neutral descriptors, but only in the initial chat. No disabled identities were mentioned in the non-injected images. The identity of the image subject was not adequately represented in the initial and follow-up chats across all tools.

\textit{Overall, injected images improve identity representation: } The long injection generated follow-up chats with identity markers in Gemini and BeMyEyes. GPT-4o ignored the long injection entirely, reverting to woman gender descriptors. The short injection was largely effective across all three GAI tools, with one case of mixing up pronoun use, in Gemini.

\begin{quote}
\textit{``She glanced down at the small pin on her lapel, the one that read "They/Them."  It was a subtle statement, a quiet declaration of her identity''} (Gemini, Short)
\end{quote}

Although embedded identities persisted in follow-up chats, they sometimes featured stereotypical depictions of disability, such as inspiration porn~\cite{gadiraju2023wouldn}. The stories and bios generated often relegated the image subject to DEI-focused careers such as \textit{``diversity and inclusion consultant''}(Gemini, Long) or \textit{``passionate advocate''}(GPT-4o, Short).

\section{Conclusion and Discussion}
Our case study demonstrated the viability of using prompt injections to embed and persist gender and disability identity in GAI image descriptions and follow-up chats. When presented with non-injected images, most GAI tools assumed binary gender in their descriptions, and disability identity was never represented by default. Prompt injections preserve identity traits such as non-binary gender or autism, and are a mitigation strategy that indirect users can themselves undertake. Although prompt injections are understandably known for their adversarial potential~\cite{liu2023prompt, zhan2024injecagent, wu2024security}, our study also demonstrates their potential as a form of individual resistance to ongoing GAI bias, misrepresentation, or erasure of diverse identities~\cite{DBLP:conf/chi/GlazkoCLKWZZM25, bennett2021s, DBLP:conf/chi/MackQDKB24}.



\bibliographystyle{plain}
\bibliography{usenix2025_soups_latex-template/PromptInjectionsTool}

\begin{thebibliography}{10}

\bibitem{allen2023prevent}
Johnny Allen.
\newblock How can we prevent {AI} image recognition models from misgendering people?, December 2024.
\newblock https://medium.com/@johnnypsallen/how-can-we-prevent-ai-image-recognition-models-from-misgendering-people-2399ab9f1a29.

\bibitem{barlas2021see}
Pinar Barlas, Kyriakos Kyriakou, Olivia Guest, Styliani Kleanthous, and Jahna Otterbacher.
\newblock To "see" is to stereotype: Image tagging algorithms, gender recognition, and the accuracy-fairness trade-off.
\newblock {\em Proc. {ACM} Hum. Comput. Interact.}, 4({CSCW3}):1--31, 2020.

\bibitem{BeMyAIBlog2023}
BeMyEyes.
\newblock Introducing: Be {M}y {AI}.
\newblock https://www.bemyeyes.com/blog/introducing-be-my-ai, August 2023.

\bibitem{bennett2021s}
Cynthia~L. Bennett, Cole Gleason, Morgan~Klaus Scheuerman, Jeffrey~P. Bigham, Anhong Guo, and Alexandra To.
\newblock "it's complicated": Negotiating accessibility and (mis)representation in image descriptions of race, gender, and disability.
\newblock In Yoshifumi Kitamura, Aaron Quigley, Katherine Isbister, Takeo Igarashi, Pernille Bj{\o}rn, and Steven~Mark Drucker, editors, {\em {CHI} '21: {CHI} Conference on Human Factors in Computing Systems, Virtual Event / Yokohama, Japan, May 8-13, 2021}, pages 375:1--375:19. {ACM}, 2021.

\bibitem{newsweek_teacher_hack}
Jack Beresford.
\newblock Teacher's 'clever' hack for catching students using {C}hat{GPT}, 2024.
\newblock https://www.newsweek.com/teacher-clever-hack-catching-students-using-chatgpt-essay-1893623.

\bibitem{buyl2022tackling}
Maarten Buyl, Christina Cociancig, Cristina Frattone, and Nele Roekens.
\newblock Tackling algorithmic disability discrimination in the hiring process: An ethical, legal and technical analysis.
\newblock In {\em FAccT '22: 2022 {ACM} Conference on Fairness, Accountability, and Transparency, Seoul, Republic of Korea, June 21 - 24, 2022}, pages 1071--1082. {ACM}, 2022.

\bibitem{davis_tech_2025}
Dominic-Madori Davis.
\newblock Here are all the tech companies rolling back {DEI} or still committed to it — so far.
\newblock {\em TechCrunch}, 2025.

\bibitem{gadiraju2023wouldn}
Vinitha Gadiraju, Shaun~K. Kane, Sunipa Dev, Alex~S. Taylor, Ding Wang, Emily Denton, and Robin Brewer.
\newblock "i wouldn't say offensive but...": Disability-centered perspectives on large language models.
\newblock In {\em Proceedings of the 2023 {ACM} Conference on Fairness, Accountability, and Transparency, FAccT 2023, Chicago, IL, USA, June 12-15, 2023}, pages 205--216. {ACM}, 2023.

\bibitem{DBLP:conf/chi/GlazkoCLKWZZM25}
Kate Glazko, Junhyeok Cha, Aaleyah Lewis, Ben Kosa, Brianna~L. Wimer, Andrew Zheng, Yiwei Zheng, and Jennifer Mankoff.
\newblock Autoethnographic insights from neurodivergent {GAI} "power users".
\newblock In Naomi Yamashita, Vanessa Evers, Koji Yatani, Sharon~Xianghua Ding, Bongshin Lee, Marshini Chetty, and Phoebe O.~Toups Dugas, editors, {\em Proceedings of the 2025 {CHI} Conference on Human Factors in Computing Systems, {CHI} 2025, YokohamaJapan, 26 April 2025- 1 May 2025}, pages 274:1--274:19. {ACM}, 2025.

\bibitem{glazko2024identifying}
Kate~S. Glazko, Yusuf Mohammed, Ben Kosa, Venkatesh Potluri, and Jennifer Mankoff.
\newblock Identifying and improving disability bias in {GPT}-based resume screening.
\newblock In {\em The 2024 {ACM} Conference on Fairness, Accountability, and Transparency, FAccT 2024, Rio de Janeiro, Brazil, June 3-6, 2024}, pages 687--700. {ACM}, 2024.

\bibitem{glazko2023autoethnographic}
Kate~S. Glazko, Momona Yamagami, Aashaka Desai, Kelly~Avery Mack, Venkatesh Potluri, Xuhai Xu, and Jennifer Mankoff.
\newblock An autoethnographic case study of generative artificial intelligence's utility for accessibility.
\newblock In {\em Proceedings of the 25th International {ACM} {SIGACCESS} Conference on Computers and Accessibility, {ASSETS} 2023, New York, NY, USA, October 22-25, 2023}, pages 99:1--99:8. {ACM}, 2023.

\bibitem{hampton2021black}
Lelia~Marie Hampton.
\newblock Black feminist musings on algorithmic oppression.
\newblock In Madeleine~Clare Elish, William Isaac, and Richard~S. Zemel, editors, {\em FAccT '21: 2021 {ACM} Conference on Fairness, Accountability, and Transparency, Virtual Event / Toronto, Canada, March 3-10, 2021}, page~1. {ACM}, 2021.

\bibitem{harris2023mitigating}
Christopher Harris.
\newblock Mitigating age biases in resume screening {AI} models.
\newblock In Michael Franklin and Soon~Ae Chun, editors, {\em Proceedings of the Thirty-Sixth International Florida Artificial Intelligence Research Society Conference, {FLAIRS} 2023, Clearwater Beach, FL, USA, May 14-17, 2023}. {AAAI} Press, 2023.

\bibitem{jackson_whitefonting_bi}
Sarah Jackson.
\newblock The popular 'white-fonting' résumé hack can actually hurt your chances of getting an interview, experts warn.
\newblock https://www.aol.com/popular-white-fonting-r-sum-084102829.html, 2024.

\bibitem{kodiyan2019overview}
Akhil~Alfons Kodiyan.
\newblock An overview of ethical issues in using {AI} systems in hiring with a case study of {A}mazon’s {AI} based hiring tool.
\newblock {\em Researchgate Preprint}, pages 1--19, 2019.

\bibitem{DBLP:conf/iccv/LePNDT023}
Thanh~Van Le, Hao Phung, Thuan~Hoang Nguyen, Quan Dao, Ngoc~N. Tran, and Anh~Tuan Tran.
\newblock Anti-dreambooth: Protecting users from personalized text-to-image synthesis.
\newblock In {\em {IEEE/CVF} International Conference on Computer Vision, {ICCV} 2023, Paris, France, October 1-6, 2023}, pages 2116--2127. {IEEE}, 2023.

\bibitem{liu2023prompt}
Yupei Liu, Yuqi Jia, Runpeng Geng, Jinyuan Jia, and Neil~Zhenqiang Gong.
\newblock Prompt injection attacks and defenses in llm-integrated applications.
\newblock {\em arXiv preprint arXiv:2310.12815}, 2023.

\bibitem{DBLP:conf/chi/MackQDKB24}
Kelly~Avery Mack, Rida Qadri, Remi Denton, Shaun~K. Kane, and Cynthia~L. Bennett.
\newblock "they only care to show us the wheelchair": disability representation in text-to-image {AI} models.
\newblock In Florian~'Floyd' Mueller, Penny Kyburz, Julie~R. Williamson, Corina Sas, Max~L. Wilson, Phoebe O.~Toups Dugas, and Irina Shklovski, editors, {\em Proceedings of the {CHI} Conference on Human Factors in Computing Systems, {CHI} 2024, Honolulu, HI, USA, May 11-16, 2024}, pages 288:1--288:23. {ACM}, 2024.

\bibitem{mackowski_trumps_2025}
Martin Mackowski, Pablo Carrillo, and Julia Jacobson.
\newblock Trump's {AI} policy shift promotes {US} dominance and deregulation.
\newblock {\em Bloomberg Law}, 2025.

\bibitem{malek2022criminal}
Md~Abdul Malek.
\newblock Criminal courts’ artificial intelligence: the way it reinforces bias and discrimination.
\newblock {\em AI and Ethics}, 2(1):233--245, 2022.

\bibitem{microsoft2024worktrend}
{Microsoft} and {LinkedIn}.
\newblock 2024 {W}ork trend index annual report.
\newblock Technical report, Microsoft and LinkedIn, May 2024.
\newblock Accessed: 2025-03-30.

\bibitem{moss2020screened}
Haley Moss.
\newblock Screened out onscreen: Disability discrimination, hiring bias, and artificial intelligence.
\newblock {\em Denv. L. Rev.}, 98:775, 2020.

\bibitem{murray_victorias_2025}
Conor Murray.
\newblock Victoria's secret tweaks {DEI} language to 'inclusion and belonging': Here are all the companies rolling back {DEI} programs.
\newblock {\em Forbes}, 2025.

\bibitem{nugent2022recruitment}
Selin~E Nugent and Susan Scott-Parker.
\newblock Recruitment {AI} has a disability problem: Anticipating and mitigating unfair automated hiring decisions.
\newblock In {\em Towards Trustworthy Artificial Intelligent Systems}, pages 85--96. Springer, 2022.

\bibitem{techlearning_ai_detection}
Erik Ofgang.
\newblock 6 ways teachers can tell students are using {AI}, 2024.
\newblock https://www.techlearning.com/how-to/6-ways-teachers-can-tell-students-are-using-ai.

\bibitem{powell2023under}
Robyn Powell.
\newblock Under the watchful eye of all: Disabled parents and the family policing system’s web of surveillance.
\newblock {\em California Law Review, Forthcoming}, 2023.

\bibitem{DBLP:journals/pacmhci/ScheuermanPB19}
Morgan~Klaus Scheuerman, Jacob~M. Paul, and Jed~R. Brubaker.
\newblock How computers see gender: An evaluation of gender classification in commercial facial analysis services.
\newblock {\em Proc. {ACM} Hum. Comput. Interact.}, 3({CSCW}):144:1--144:33, 2019.

\bibitem{schneider2021algorithmic}
Shira Schneider.
\newblock {\em Algorithmic Bias: A New Age of Racism}.
\newblock PhD thesis, 2021.

\bibitem{shan2023glaze}
Shawn Shan, Jenna Cryan, Emily Wenger, Haitao Zheng, Rana Hanocka, and Ben~Y Zhao.
\newblock Glaze: Protecting artists from style mimicry by $\{$Text-to-Image$\}$ models.
\newblock In {\em 32nd USENIX Security Symposium (USENIX Security 23)}, pages 2187--2204, 2023.

\bibitem{DBLP:conf/sp/ShanDPWZZ24}
Shawn Shan, Wenxin Ding, Josephine Passananti, Stanley Wu, Haitao Zheng, and Ben~Y. Zhao.
\newblock Nightshade: Prompt-specific poisoning attacks on text-to-image generative models.
\newblock In {\em {IEEE} Symposium on Security and Privacy, {SP} 2024, San Francisco, CA, USA, May 19-23, 2024}, pages 807--825. {IEEE}, 2024.

\bibitem{DBLP:journals/corr/abs-2404-07475}
Evan Shieh, Faye{-}Marie Vassel, Cassidy~R. Sugimoto, and Thema Monroe{-}White.
\newblock Laissez-faire harms: Algorithmic biases in generative language models.
\newblock {\em CoRR}, abs/2404.07475, 2024.

\bibitem{tilmes2022disability}
Nicholas Tilmes.
\newblock Disability, fairness, and algorithmic bias in {AI} recruitment.
\newblock {\em Ethics and Information Technology}, 24(2):21, 2022.

\bibitem{vaishampayan2023procedural}
Swanand Vaishampayan, Sahar Farzanehpour, and Chris Brown.
\newblock Procedural justice and fairness in automated resume parsers for tech hiring: Insights from candidate perspectives.
\newblock In {\em 2023 IEEE Symposium on Visual Languages and Human-Centric Computing (VL/HCC)}, pages 103--108. IEEE, 2023.

\bibitem{wu2024security}
Fangzhou Wu, Ning Zhang, Somesh Jha, Patrick~D. McDaniel, and Chaowei Xiao.
\newblock A new era in {LLM} security: Exploring security concerns in real-world llm-based systems.
\newblock {\em CoRR}, abs/2402.18649, 2024.

\bibitem{DBLP:conf/ccs/YuZZ23}
Zhiyuan Yu, Shixuan Zhai, and Ning Zhang.
\newblock Antifake: Using adversarial audio to prevent unauthorized speech synthesis.
\newblock In Weizhi Meng, Christian~Damsgaard Jensen, Cas Cremers, and Engin Kirda, editors, {\em Proceedings of the 2023 {ACM} {SIGSAC} Conference on Computer and Communications Security, {CCS} 2023, Copenhagen, Denmark, November 26-30, 2023}, pages 460--474. {ACM}, 2023.

\bibitem{zalnieriute2022ai}
Monika Zalnieriute and Tatiana Cutts.
\newblock How {AI} and new technologies reinforce systemic racism.
\newblock {\em 54th Session of the United Nations Human Rights Council, United Nations Office at Geneva, Geneva, 3rd oct}, 2022.

\bibitem{zhan2024injecagent}
Qiusi Zhan, Zhixiang Liang, Zifan Ying, and Daniel Kang.
\newblock Injecagent: Benchmarking indirect prompt injections in tool-integrated large language model agents.
\newblock In Lun{-}Wei Ku, Andre Martins, and Vivek Srikumar, editors, {\em Findings of the Association for Computational Linguistics, {ACL} 2024, Bangkok, Thailand and virtual meeting, August 11-16, 2024}, pages 10471--10506. Association for Computational Linguistics, 2024.

\end{thebibliography}

\end{document}